\documentstyle[preprint,aps]{revtex}
\begin{document}
\title{\text{Theory of exchange coupling in disordered magnetic multilayers}}
\author{A.Yu. Zyuzin}
\address{A.F.Ioffe Physical-Technical Institute RAS, 194021 St.Petersburg, \\
Russia}
\maketitle

\begin{abstract}
We consider mechamism of exchange coupling based on interaction between
electrons in nonmagnetic layer. Depending on ratio of inverse time of
diffusion of electrons between ferromagnetic layers and ferromagnetic
splitting of conducting electrons this mechanism describes transition from
ferromagnetic to concollinear ordering of magnetizations of ferromagnetic
layers.
\end{abstract}

\section{Introduction and main results.}

In metallic ferromagnet-nonferromagnet-ferromagnet multilayers (see fig. 1)
magnetic structure oscillates between ferromagnetic and antiferromagnetic
orientations of the ferromagnets' magnetizations as a function of thickness
of nonmagnetic metal $L$ with a period of order of the Fermi wave length 
\cite{bai}\cite{par} \cite{levy}\cite{yang}\cite{hein}\cite{yaf}. The
explanation of this phenomenon is based on the fact that the interlayer
coupling is due to Ruderman-Kittel interaction between electron spins in
different ferromagnets.

Further investigations discovered structures with perpendicular orientations
of the ferromagnets' magnetizations (see for review \cite{slon}). Often
phenomenological coupling between magnetizations of ferromagnetic layers can
be represented as sum of bilinear and biquadratic contributions 
\begin{equation}
E\left( \varphi \right) =J_{1}\cos \varphi +J_{2}\cos ^{2}\varphi
\end{equation}

Here $\varphi $ is angle between directions of magnetizations of
ferromagnetic films. Bilinear constant $J_{1}$ oscillates as function of
interlayer distance $L.$\ In case of large positive biquadratic constant $%
J_{2}$ minimum of $E\left( \varphi \right) $ corresponds to $\varphi =\pi /2$
. As explained by Slonzevskii large positive biquadratic coupling might be
result of spatial fluctuations of bilinear coupling $J_{1}$ due to
ferromagnet-nonferromagnet surface roughness \cite{slon}.

In disordered system, when $L$ is larger than electron mean free path $l$ ,
RKKY interaction $\left\langle J_{1}\right\rangle $, averaged over
realizations of scattering potential exponentially decreases \cite{deGen}.
At the same time fluctuations of local exchange become much larger than $%
\left\langle J_{1}\right\rangle $ \cite{RKKY} giving rise to biquadratic
contribution $J_{2}>>\left| \left\langle J_{1}\right\rangle \right| $ \cite
{spiv}.

Here we propose mechanism of coupling in disordered multilayers based on
interaction between electrons in nonmagnetic layer. Spin fluctuations in
system of interacting electrons give rise to contribution to thermodynamic
potential\cite{alt}, which depends on magnetic field or, in our case, on
relative orientation of magnetizations in ferromagnetic layers. Here we show
that in magnetic multilayer this mechanism describes transition between
ferromagnetic and noncolinear ordering with increasing distance between
ferromagnetic layers or value of ferromagnetic splitting of conducting
electrons.

We assume that magnetic multilayer can be described by Hamiltonian

\begin{equation}
H=H_{0}+\epsilon _{exc}\int\limits_{F}d{\bf r}\Psi _{\alpha }^{+}\left( {\bf %
r}\right) {\bf n}\left( z\right) {\bf \sigma }_{\alpha \beta }\Psi _{\beta
}\left( {\bf r}\right) +H_{int}
\end{equation}

Here $H_{0}$ is Hamiltonian of free electrons in random field. Second term
describes exchange field in ferromagnetic layers. $\epsilon _{exc}$ is
ferromagnetic splitting of conducting electrons. ${\bf n}\left( z\right) $
is unit vector of direction of magnetization of ferromagnetic layers. ${\bf n%
}\left( z\right) ={\bf n}_{1}\ $at $z<-L/2$ and ${\bf n}\left( z\right) =%
{\bf n}_{2}\ $at $z>L/2$ as it is shown on figure 1. $\Psi _{\alpha
}^{+}\left( {\bf r}\right) $ and $\Psi _{\beta }\left( {\bf r}\right) $ are
creation and annihilation operators, ${\bf \sigma }_{\alpha \beta }$ are
Pauli matrixes. Integration in second term is over ferromagnetic layers. The
last term $H_{int}$ describes Coulomb interaction between electrons in
nonmagnetic layer. We assume that interaction in ferromagnetic layers is
taken into account self consistently in $\epsilon _{exc}$.

Details of calculation are given in the last part of the paper. Here we
present the main results. Characteristic energies in the problem are
ferromagnetic splitting of conducting electrons $\epsilon _{exc}$ and
Thouless energy $E_{c}\equiv D/L^{2}$ . $D$ is diffusion constant of
conduction electrons. We assume that it is the same in nonmagnetic and
ferromagnetic layers.

In case of small thickness when $E_{c}>\epsilon _{exc}$ coupling between
ferromagnetic layers has bilinear form and coupling energy per unit area is

\begin{equation}
E\left( \varphi \right) =-\frac{F}{8\left( 4\pi L\right) ^{2}}\frac{\epsilon
_{exc}^{2}}{E_{c}}\cos \varphi
\end{equation}

Here $F$ is characteristic constant of interaction in diffusion channel\cite
{alt}. It is positive for Coulomb repulsion between electrons. Let us note
that in this regime coupling (3) does not depend on $L$ . Minimum of (3)
corresponds to ferromagnetic orientation of magnetizations in multilayer $%
\varphi =0$. Note that result is obtained in limit when $L>l$ , or $\epsilon
_{exc}$ smaller that inverse mean free time $D/l^{2}$.

At larger distance $L$ when $E_{c}<\epsilon _{exc}$ coupling has biquadratic
form and coupling energy per unit area is 
\begin{equation}
E\left( \varphi \right) \simeq \frac{F}{\left( 4\pi L\right) ^{2}}E_{c}\cos
^{2}\varphi
\end{equation}

This quantity decreases as $L^{-4}$ with increasing distance. Minimum of
coupling energy corresponds to noncollinear state $\varphi =\pi /2$.

Both expressions are given for the case of infinite thickness of
ferromagnetic layers. Calculation show that in case $d>\sqrt{D/\epsilon
_{exc}}$ coupling weakly depends on $d$.

Results (3) and (4) are also valid provided $L<\sqrt{D/T}$ . In opposite
case coupling energy decreases exponentially as $\exp \left( -\sqrt{%
%TCIMACRO{\dfrac{8\pi T}{D}}%
%BeginExpansion
{\displaystyle{8\pi T \over D}}%
%EndExpansion
}L\right) $ .

Let us compare results (2) and (3) with biquadratic contribution due to
mesoscopic fluctuations of RKKY interaction \cite{spiv}, which is $J_{2}\sim 
\frac{1}{L^{2}}\frac{E_{c}^{2}}{Ad}$ at $E_{c}<\epsilon _{exc}$ and $%
J_{2}\sim \frac{1}{L^{2}}\frac{\epsilon _{exc}^{3}}{AdE_{c}}$ at $%
E_{c}>\epsilon _{exc}$. Here $A$ is an intralayer ferromagnetic stiffness
and thickness $d>\sqrt{D/\epsilon _{exc}}$ . \ 

The quantity $J_{2}$ decreases with $L$ much faster than (4). Also for $%
\epsilon _{exc}/Ad<<1$ , $F\approx 1$ coupling energy given by expressions
(3) and (4) is larger than biquadratic contribution due to mesoscopic
fluctuations of RKKY in whole range of distances. In this case with
increasing distance $L>>l$ system is undergo transition between
ferromagnetic and perpendicular $\varphi =\pi /2$ ordering. Such transition
was observed in \cite{exper}.

\section{Derivation of results.}

Correction to thermodynamic potential which depends on $\epsilon _{exc}{\bf n%
}\left( z\right) $ is given by expression \cite{alt}

\begin{equation}
\delta \Omega =\frac{F}{4}T\sum\limits_{\left| \omega _{n}\tau \right|
<1;\alpha ,\beta }\left| \omega _{n}\right| \int \frac{d^{2}{\bf q}}{\left(
2\pi \right) ^{2}}\int\limits_{\left| z\right| <L/2}dzD_{\beta \beta
}^{\alpha \alpha }\left( z,z,{\bf q},\omega _{n}\right)
\end{equation}

Here constant $F$ describes screened Coulomb interaction in diffusion
channel. $\omega _{n}=2\pi nT$ is Matsubara frequency. $\tau $ is electron
mean free time.

Diffusion ladder satisfies equation

\begin{equation}
\begin{array}{c}
\left( -D\frac{d^{2}}{dz^{2}}+Dq^{2}+\left| \omega _{n}\right| \right)
D_{\mu \eta }^{\alpha \beta }+i\epsilon _{exc}{\bf n}\left( z\right) \left( 
{\bf \sigma }_{\alpha \gamma }D_{\mu \eta }^{\gamma \beta }-D_{\mu \gamma
}^{\alpha \beta }{\bf \sigma }_{\gamma \eta }\right) sign\omega _{n}= \\ 
=\delta \left( z-z^{\prime }\right) \delta _{\alpha \beta }\delta _{\mu \eta
}
\end{array}
\end{equation}

It is convenient to present solution of equation (6) at $\left| z\right|
<L/2 $ in the form

\begin{equation}
D_{\mu \eta }^{\alpha \beta }=A_{\mu \eta }^{\alpha \beta }\exp \left(
-Qz\right) +U_{\alpha \gamma }^{+}C_{\mu \gamma }^{\gamma \beta }U_{\gamma
\eta }\exp \left( Qz\right) +%
%TCIMACRO{\dfrac{\exp \left( -Q\left| z-z^{\prime }\right| \right) }{2DQ}}%
%BeginExpansion
{\displaystyle{\exp \left( -Q\left| z-z^{\prime }\right| \right)  \over 2DQ}}%
%EndExpansion
\delta _{\alpha \beta }\delta _{\mu \eta }
\end{equation}
Here we introduce $Q=\sqrt{q^{2}+\frac{\left| \omega _{n}\right| }{D}}$. $U$
\ is matrix of relative rotation of magnetizations of ferromagnetic layers.
In case when direction of magnetization in ferromagnetic layer $z<-L/2$ is
directed along $z$ axes ${\bf n}\left( z\right) =\left( 0,0,1\right) $ and
at $z>L/2$ $\ $direction is ${\bf n}\left( z\right) =\left( \sin \varphi
,0,\cos \varphi \right) $, it is matrix of rotation along $y$ axes $U=\exp
\left( 
%TCIMACRO{\dfrac{i\varphi }{2}}%
%BeginExpansion
{\displaystyle{i\varphi  \over 2}}%
%EndExpansion
\sigma _{y}\right) $.

For simplicity we consider limit of semiinfinite ferromagnetic layers. More
detail consideration shows that at $d>\sqrt{D/\epsilon _{exc}}$ results
weakly depend on thickness of ferromagnetic layers. It is convenient to
introduce boundary conditions for diffusion ladder at
ferromagnet-nonferromagnet interfaces \ taking into account that according
to equation (6) in coordinate system where spins are directed along
magnetization, components of ladder with antiparallel spins decreases as $%
\exp \left( -Q_{1}\left| z\right| \right) $ and $\exp \left( -Q_{1}^{\ast
}\left| z\right| \right) $ at $\left| z\right| >L/2$, where $Q_{1}=\sqrt{%
q^{2}+%
%TCIMACRO{\dfrac{\omega +i\epsilon _{exc}}{D}}%
%BeginExpansion
{\displaystyle{\omega +i\epsilon _{exc} \over D}}%
%EndExpansion
}$ . Components of ladder with parallel spins decreases as $\exp \left(
-Q\left| z\right| \right) $ at $\left| z\right| >L/2$. At $z=-L/2$ where $%
{\bf n}\left( z\right) =\left( 0,0,1\right) $ boundary conditions are

\begin{equation}
\begin{array}{c}
\left( 
%TCIMACRO{\dfrac{d}{dz}}%
%BeginExpansion
{\displaystyle{d \over dz}}%
%EndExpansion
-Q_{1}\right) P_{\alpha \gamma }^{+}D_{\mu \gamma }^{\gamma \beta }P_{\gamma
\eta }^{-}=\left( 
%TCIMACRO{\dfrac{d}{dz}}%
%BeginExpansion
{\displaystyle{d \over dz}}%
%EndExpansion
-Q_{1}^{\ast }\right) P_{\alpha \gamma }^{-}D_{\mu \gamma }^{\gamma \beta
}P_{\gamma \eta }^{+}=0 \\ 
\left( 
%TCIMACRO{\dfrac{d}{dz}}%
%BeginExpansion
{\displaystyle{d \over dz}}%
%EndExpansion
-Q\right) P_{\alpha \gamma }^{\pm }D_{\mu \gamma }^{\gamma \beta }P_{\gamma
\eta }^{\pm }=0
\end{array}
\end{equation}

Here we introduce projectors of spins on $z$-axes $P_{\pm }=%
%TCIMACRO{\dfrac{1\pm \sigma _{z}}{2}}%
%BeginExpansion
{\displaystyle{1\pm \sigma _{z} \over 2}}%
%EndExpansion
$.

The same kind of boundary conditions can be introduced for rotated diffusion
ladder $U_{\alpha \gamma }D_{\mu \gamma }^{\gamma \beta }U_{\gamma \eta
}^{+} $ at $z=L/2$ . Solving system of equations (6,8) we obtain

\begin{equation}
\begin{array}{c}
\delta \Omega =-\frac{F}{2}T\sum\limits_{\left| \omega _{n}\tau \right|
<1}\left| \omega _{n}\right| \int \frac{d^{2}{\bf q}}{\left( 2\pi \right)
^{2}}\frac{L}{DQ}\times \\ 
\times \frac{\left[ \left( \left| \Lambda \right| ^{2}-\left( 
%TCIMACRO{\func{Re}}%
%BeginExpansion
\mathop{\rm Re}%
%EndExpansion
\Lambda \right) ^{2}\right) \cos \varphi +\left( \left| \Lambda \right| ^{4}-%
\frac{1}{2}\left( 
%TCIMACRO{\func{Re}}%
%BeginExpansion
\mathop{\rm Re}%
%EndExpansion
\Lambda \right) ^{2}\right) \cos ^{2}\varphi -\frac{1}{2}\left( 
%TCIMACRO{\func{Re}}%
%BeginExpansion
\mathop{\rm Re}%
%EndExpansion
\Lambda \right) ^{2}+%
%TCIMACRO{\func{Re}}%
%BeginExpansion
\mathop{\rm Re}%
%EndExpansion
\Lambda \left( 1-\frac{1}{2}\left| \Lambda \right| ^{2}-\frac{1}{2}\left|
\Lambda \right| ^{2}\cos ^{2}\varphi \right) \frac{\sinh QL}{QL}\right] }{%
\left[ 1-\left( 
%TCIMACRO{\func{Re}}%
%BeginExpansion
\mathop{\rm Re}%
%EndExpansion
\Lambda \right) ^{2}+2\left( \left| \Lambda \right| ^{2}-\left( 
%TCIMACRO{\func{Re}}%
%BeginExpansion
\mathop{\rm Re}%
%EndExpansion
\Lambda \right) ^{2}\right) \cos \varphi +\left( \left| \Lambda \right|
^{4}-\left( 
%TCIMACRO{\func{Re}}%
%BeginExpansion
\mathop{\rm Re}%
%EndExpansion
\Lambda \right) ^{2}\right) \cos ^{2}\varphi \right] }
\end{array}
\end{equation}

Here $\Lambda =%
%TCIMACRO{\dfrac{\left( Q_{1}-Q\right) }{\left( Q_{1}+Q\right) }}%
%BeginExpansion
{\displaystyle{\left( Q_{1}-Q\right)  \over \left( Q_{1}+Q\right) }}%
%EndExpansion
\exp \left( -QL\right) $. Expression (9) contains divergent terms, which do
not depend $\varphi $ and must be subtracted.

In limit of large exchange splitting when $\left| Q_{1}\right| >Q$ parameter 
$\Lambda =\exp \left( -QL\right) $ is real. In this case energy is function
of $\cos ^{2}\varphi $. Subtracting in expression (9) terms which do no
depend on angle we obtain

\begin{equation}
\delta \Omega \left( \varphi \right) =\frac{F}{4}T\sum\limits_{\left| \omega
_{n}\tau \right| <1;\alpha ,\beta }\left| \omega _{n}\right| \int \frac{d^{2}%
{\bf q}}{\left( 2\pi \right) ^{2}}\frac{L\left( 1-\Lambda \frac{\sinh QL}{QL}%
\right) }{DQ}\frac{\Lambda ^{2}\cos ^{2}\varphi }{\left( 1-\Lambda ^{2}\cos
^{2}\varphi \right) }
\end{equation}
Main contribution in expression (10) is from region where $\Lambda <1$ ,
denominator therefore gives only small correction. Neglecting it we obtain
expression (4).

In opposite limit of small exchange splitting $%
%TCIMACRO{\func{Re}}%
%BeginExpansion
\mathop{\rm Re}%
%EndExpansion
\Lambda \sim \epsilon _{ex}^{2}$ , $\left| \Lambda \right| \sim \epsilon
_{ex}$ and to the order $\epsilon _{ex}^{2}$ coupling energy is proportional
to $\cos \varphi .$

\begin{equation}
\delta \Omega \left( \varphi \right) =-\frac{F}{2}T\sum\limits_{\left|
\omega _{n}\tau \right| <1;\alpha ,\beta }\left| \omega _{n}\right| \int 
\frac{d^{2}{\bf q}}{\left( 2\pi \right) ^{2}}\frac{L\left| \Lambda \right|
^{2}}{DQ}\cos \varphi
\end{equation}
Calculating (11) at zero temperature we obtain (3). Transition between
limits (10) and (11) occurs at $\epsilon _{ex}\sim D/L^{2}$.

\section{Acknowledgement}

This work is supported by Russian Fund for Fundamental Research grant number
01-02-17794.~


\begin{references}
\bibitem{bai}  M.Baibich at all., Phys.Rev.Lett.61, 2472, 1988

\bibitem{par}  S.S.P.Parkin, N.More, K.P.Roche, Phys.Rev.Lett.64, 2304,
1990; ibid 66, 2152, 1991.

\bibitem{levy}  P.Levy, S.Zang, A.Fert, Phys.Rev.Lett.65, 1643, 1990.

\bibitem{yang}  Q.Yang at all, Phys.Rev.Lett.72, 3274, 1994.

\bibitem{hein}  B.Heinrich, J.F.Cochran, Adv. in Physics, 42, 523, 1993.

\bibitem{yaf}  Y.Yafet J.Appl.Phys. 61, 4058, 1997.

\bibitem{slon}  J.C. Slonczewski, J.Appl.Phys.73, 5957, 1993,
J.Magn.Magn.Mater. 150, 13, 1995

\bibitem{deGen}  P.de Gennes, J.Phys.Rad. 23, 230, 1962.

\bibitem{RKKY}  A.Yu. Zyuzin, B. Spivak, Sov.Phys.JETP Lett., 43, 234, 1986.

\bibitem{spiv}  A.Yu. Zyuzin, B.Z. Spivak, I. Vagner, P. Wyder, Phys.Rev.
B., 62, 13899, 2000

\bibitem{alt}  B.L.Altshuler, A.G. Aronov, A.Yu. Zyuzin, Sov.Phys.JETP 57,
889, 1983

\bibitem{exper}  P. Fuchs, U. Ramsperger, A. Vaterlaus and M. Landolt, Phys.
Rev. B 55, 12546, 1997.
\end{references}
\end{document}